# Beyond Data Scarcity: Optimizing R3GAN for Medical Image Generation from Small Datasets

Tsung-Wei Pan
Department of Electrical Engineering
National Taiwan Ocean University
Keelung 202301, Taiwan
well90111829@gmail.com

Chang-Hong Wu
Department of Electrical Engineering
National Taiwan Ocean University
Keelung 202301, Taiwan
s0979979782@gmail.com

Jung-Hua Wang*
AI Research Center
National Taiwan Ocean University
Keelung 202301
jhwang@email.ntou.edu.tw

Ming-Jer Chen
Department of Obstetrics, Gynecology
and Woman's Health, TVGH, 407219
Taichung
mingjerchen@gmail.com

Yu-Chiao Yi
Department of Obstetrics, Gynecology
and Woman's Health, TVGH, 407219
Taichung
yuchiaoyi@gmail.com

Tsung-Hsien Lee
Department of Obstetrics and
Gynecology, Chung Shan Medical
University Hospital, Taichung 40201,
Taiwan
jackth.lee@gmail.com

*Abstract*—Medical image datasets frequently exhibit significant class imbalance, a challenge that is further amplified by the inherently limited sample sizes that characterize clinical imaging data. Using human embryo time-lapse imaging (TLI) as a case study, this work investigates how generative adversarial networks (GANs) can be optimized for small datasets to generate realistic and diagnostically meaningful images. Based on systematic experiments with R3GAN, we established effective training strategies and designed an optimized configuration for 256×256-resolution datasets, featuring a full burn-in phase and a low, gradually increasing $\gamma$ range (5 → 40). The generated samples were used to balance an imbalanced embryo dataset, leading to substantial improvement in classification performance. The recall and F1-score of *t3* increased from 0.06 to 0.69 and 0.11 to 0.60, respectively, without compromising other classes. These results demonstrate that tailored R3GAN training strategies can effectively alleviate data scarcity and improve model robustness in small-scale medical imaging tasks.

## I. Introduction

In recent years, machine learning (ML) techniques have demonstrated significant promise in the medical domain, particularly for clinical diagnosis and decision-making through medical image analysis. These data-driven approaches leverage advancements in computer vision to enhance the consistency and reliability of clinical assessments. Despite this potential, the development of robust medical ML models faces a critical obstacle: data scarcity and heterogeneity. Unlike large-scale natural image datasets (e.g., ImageNet), medical data acquisition is severely constrained by ethical and privacy regulations (e.g., HIPAA [1] and GDPR [2]), limiting large-scale sharing. Furthermore, expert annotation is costly and time-consuming, and certain conditions are characterized by biological rarity and temporal specificity, which further restrict the availability of representative samples. This inherent data limitation poses a fundamental challenge to the stability and performance of deep learning models.

A representative example of this challenge lies in the Time-Lapse Imaging (TLI) of human embryos. In reproductive medicine, one of the core objectives is to quantify the Absolute Cleavage Timing (ACT) based on TLI sequences. This process necessitates a model that can reliably classify embryo images into their respective developmental stages, where each class transition marks a distinct cleavage event in time. However, certain intermediate stages are inherently transient and temporally specific. For instance, the three-cell (t3) stage, which occurs between the rapid two-cell and four-cell divisions, lasts only a very short and variable duration. This biological constraint results in a significant scarcity of high-quality t3 samples [3]. Coupled with the stringent privacy regulations of reproductive medicine, TLI data thus serves as a prototypical example of a small, imbalanced medical image dataset.

To address this problem, we first investigated a range of conventional data augmentation techniques—such as random scaling, translation, rotation, and horizontal flipping—that are widely adopted to improve model robustness against spatial and photometric variations [4]–[6]. In principle, these operations increase the diversity of image appearances by introducing geometric transformations, thereby helping models learn invariance to minor positional shifts or illumination changes. However, such augmentations merely generate local perturbations within the vicinity of the existing data manifold and thus cannot effectively expand the intrinsic distribution of embryo morphology. For embryonic TLI datasets, this limitation is even more pronounced due to the unique biological and imaging characteristics of embryos. Embryo images typically exhibit near-perfect circular morphology and strong rotational symmetry; in other words, operations such as flipping or rotation provide almost no new structural variation [7]. Instead of enhancing data diversity, these transformations often yield redundant samples that overlap heavily in feature space. Additionally, the overall grayscale contrast and centralized framing of TLI images make photometric or positional augmentations (e.g., brightness jitter, random cropping) less impactful compared to natural image tasks. Mixing-based augmentation techniques, such as MixUp [8] and CutMix [9], further exacerbate this problem in biomedical contexts. By linearly combining or spatially merging multiple embryo images, these methods can produce anatomically implausible composites that disrupt the visual integrity of blastomere boundaries, alter cell texture continuity, and compromise biological realism.

Collectively, these observations suggest that while traditional augmentation can enhance a model's robustness to imaging noise and minor perturbations, it is inherently constrained by its inability to introduce new structural variations or alleviate class-level data imbalance. To overcome the aforementioned limitations of conventional augmentation and further alleviate the issue of limited data availability, we adopt a generative modeling strategy for medical image augmentation. Generative Adversarial Networks (GANs) [10]–[14] have shown remarkable capability in learning underlying visual distributions and synthesizing realistic images. However, while GANs produce stable and high-quality results on large-scale datasets, their training becomes unstable and less effective under limited data conditions, as in the case of TLI datasets. A GAN consists of a generator that transforms Gaussian noise into images and a discriminator that guides the generator's learning by distinguishing real from synthetic samples. This adversarial setup enables the model to learn complex mappings from a simple, well-sampled latent distribution to the intricate, data-scarce embryo domain.

## II. RELATED WORK

### A. Architecture Centric Method

In [3], the authors tackle data imbalance through an architecture-centered approach rather than conventional resampling or augmentation. Their framework integrates multiple task-specific components, including several ResNet-based [15] classifiers, an instance segmentation network, and an Expectation–Maximization (EM) module. Each classifier focuses on recognizing a particular cleavage stage, while the segmentation model extracts spatial features such as blastomere area and contour. The EM module then statistically refines the predicted cleavage timings by aligning them with patient-specific temporal distributions, thereby compensating for uneven representation across developmental stages. This multi-module design effectively distributes learning responsibilities and enhances the model's ability to capture diverse embryo characteristics, even under imbalanced data conditions. Nevertheless, such an architecture inevitably increases model complexity, parameter interactions, and computational cost. When data availability is limited, this added complexity may introduce risks such as slower convergence or potential overfitting, underscoring the need to balance architectural sophistication with dataset scale. To mitigate these concerns, we adopt a data-centric perspective—focusing on improving the diversity, representativeness, and distributional balance of the training data itself—so that model performance can be enhanced without further increasing structural complexity.

### B. R3GAN vs. GANs

Traditional GANs often encounter critical challenges such as unstable training, non-convergence, and mode collapse. Because GANs are formulated as a minimax game between a generator and a discriminator, their gradients frequently oscillate or diverge, making it difficult to reach equilibrium. The discriminator in standard GANs separates real and fake samples independently, which can cause the generator to collapse to a few limited modes of the data distribution, reducing diversity and overall image quality.

R3GAN [16] addresses these issues by extending the Relativistic Pairing GAN (RpGAN) [17] framework with zero-centered gradient penalties (R1 and R2) [18], [19]. This regularization constrains the discriminator's gradients around zero, suppressing gradient explosion and ensuring more stable adversarial optimization. The pairwise RpGAN loss further compares real and fake samples relatively, allowing each real sample to define a local decision boundary. This design prevents mode collapse and encourages the generator to learn multiple modes of the data distribution, leading to better diversity and convergence. Through this theoretically grounded loss formulation, R3GAN eliminates the need for empirical heuristics such as normalization or weight-initialization tricks while achieving superior fidelity and stability. Owing to these advantages, R3GAN has emerged as one of the most robust and effective GAN variants, and we therefore adopt it as the core generative model in this study for embryo TLI image synthesis under limited data conditions.

### C. R3GAN Training Configuration and Dynamics on Large-Scale Datasets

Before discussing the small-scale training behavior in the next chapter, it is necessary to clarify several fundamental conventions and hyperparameters in R3GAN. The unit **kimg** refers to one thousand images processed during training, including both real and generated samples. This convention allows progress to be expressed independently of batch size or dataset scale, providing a consistent measure across different experiments.

A critical concept in R3GAN's training design is the **burn-in phase**, an early stage during which all major hyperparameters—learning rate, R1/R2 regularization strength ($\gamma$), Adam $\beta_2$, EMA half-life, and augmentation probability—are gradually transitioned from their initial to final values following a cosine schedule. The purpose of burn-in is to allow the generator and discriminator to reach a balanced state before introducing full regularization. A large initial $\gamma$ helps smooth both model distributions early in training, while the optimizer's low initial $\beta_2$ enables rapid adaptation to changing gradient magnitudes. Meanwhile, augmentation is disabled at the start and progressively enabled only when the discriminator begins to overfit. For large-scale datasets such as FFHQ 256×256 and ImageNet, R3GAN employs a short burn-in duration (8–20% of total training) and a rapid cosine decay of $\gamma$ (e.g., 150 → 15). These choices rely on the statistical stability of massive datasets, where diverse and abundant samples prevent overfitting and allow the discriminator to learn stable boundaries early. Consequently, aggressive hyperparameter decay accelerates convergence and reduces computational cost.

During large-scale experiments, R3GAN demonstrates fast initial convergence followed by a stable equilibrium, as evidenced by rapidly decreasing Fréchet Inception Distance (FID) [20] scores within the first few hundred kimg and steady improvement thereafter. The relativistic loss and zero-centered gradient penalties effectively suppress gradient explosion and mode collapse, while the cosine-scheduled burn-in ensures smooth transitions between early exploration and late refinement. Overall, the large-scale training configuration establishes the foundation from which this study explores alternative strategies for **small datasets**, where limited diversity and high gradient variance fundamentally alter these dynamics.

## III. METHOD: TRAINING STRATEGY

This section describes the progressive development of a stable training strategy for R3GAN on the small-scale human embryo time-lapse imaging (TLI) dataset provided by Taichung Veterans General Hospital (TCVGH), comprising approximately 3.6k images. The goal was to adapt the original large-dataset configuration of R3GAN to a data-constrained environment, where conventional training led to high FID values and unstable loss dynamics. Through a series of controlled experiments (exps 003–017), our investigation primarily focused on the **burn-in phase** and the **$\gamma$ parameter**

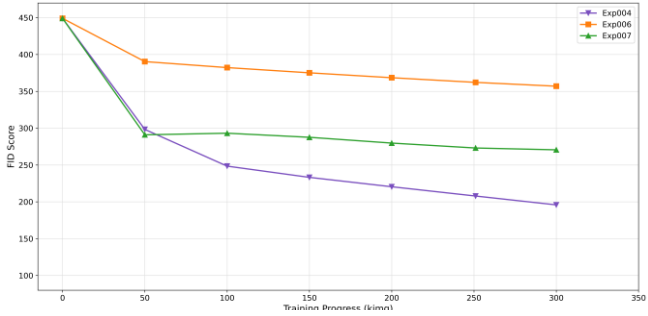

Fig. 1. FID Score comparison of three kinds of burn-in strategy.

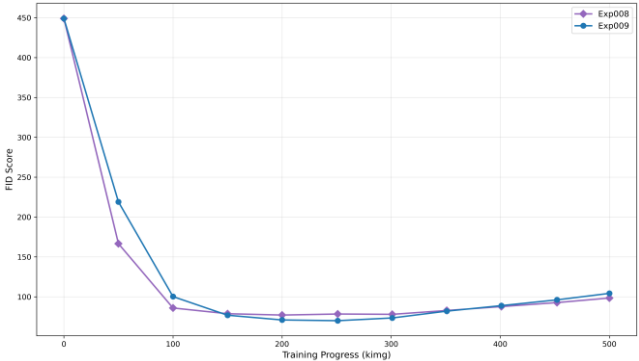

Fig. 3. FID comparison between exp008 and exp009.

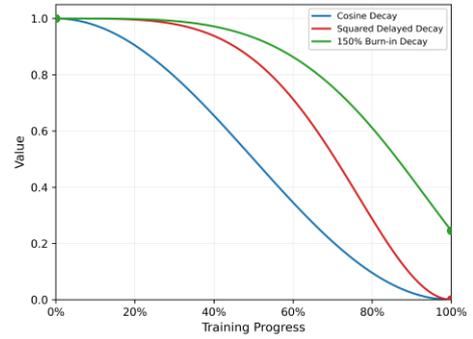

Fig. 2. Three kinds of decay strategy (assuming the parameter starts at 1 and decays to 0).

**controlling the R1 and R2 regularizations**, as these components are central to the R3GAN design.

Only representative experiments (exps 003–017) are presented bel ow, while intermediate trials with overlapping settings are omitted for clarity. Note that the experiments are not necessarily discussed in sequential order, as each subsection focuses on the configurations that led to distinct observations.

*A. Original Baseline and Burn-in Exploration*

Because the embryo dataset in this study most closely matches the FFHQ-256×256 setting in the original R3GAN paper, all images were resized to 256×256 and we first adopted a configuration similar to FFHQ-256 to enable fair comparison. To accommodate the small-dataset regime, the total training length was fixed at **300 kimg**, and we systematically varied the burn-in duration.

In the original R3GAN experiments, burn-in typically spans 8–20% of total training. Following this, **exp 006** used a 20% burn-in, but performance was poor, with the best FID remaining above **250**. We then tested longer warm-ups: **exp 007** with 50% burn-in and **exp 004** with 100% burn-in (Fig. 1). Both outperformed the 20% baseline, and the 100% configuration (exp 004) was better than the 50% configuration (exp 007).

These results indicate that short burn-in schedules are unsuitable for small datasets. Since a small dataset provides limited learning signal, an overly short burn-in unduly compresses the model's effective learning time and pushes training prematurely into a late-stage, low-update regime, leading to under-training. Therefore, subsequent experiments adopted the 100% burn-in strategy.

*B. Delayed-Decay Scheduling*

In earlier experiments, **exp 003** directly followed the FFHQ-256 training configuration from the original R3GAN paper, except that the total training length was shortened to **300 kimg**. Because its burn-in duration still corresponded to the original **2 Mimg** schedule used for large datasets, the hyperparameters effectively remained constant throughout training without any decay. Interestingly, this configuration achieved a rapid initial FID drop to **106**, outperforming exps 004, 006, and 007, although the FID later rebounded during the final stage.

Comparing both patterns revealed a clear trade-off: maintaining high hyperparameter values (i.e., little or no decay) allows a deeper initial FID drop but leads to later degradation, while enforcing a smooth, continuously decaying schedule ensures stable convergence but limits early improvement. To combine the advantages of both behaviors, we introduced a delayed-decay schedule—a curve that retains high hyperparameter values during the early phase and gradually decays toward the end. The cosine decay function used in R3GAN is given by $0.5(1 + \cos(\pi x))$ (as shown by the blue line in Fig. 2), and by squaring the x-axis term, a "flattened head, steep tail" decay can be produced (red line in Fig. 2). An alternative delayed-decay strategy is to extend the burn-in phase to 150%, which similarly delays the onset of decay (green line in Fig. 2).

These two approaches correspond to **exp 008** (squared-decay schedule) and **exp 009** (150% burn-in). Exp 008 achieved a major improvement with a best FID of **77**, showing smoother late-stage behavior than exp 003. Exp 009 further reduced the FID to **70**, slightly outperforming 008. However, when comparing the two curves (Fig. 3), exp 009 showed weaker early and late performance, with only a marginal mid-phase advantage.

To verify that this difference was not due to random variation, two additional trials with different random seeds were performed (**exp 010** and **011**). The overall pattern remained consistent—while some fluctuation existed, the 150% burn-in strategy (011) still outperformed the squared-decay schedule (010). We therefore conclude that retaining stronger hyperparameter values in the early phase benefits small-dataset training, with extended burn-in offering a more reliable implementation of this principle.

*C. Reversing the γ Schedule*

In the original FFHQ-256 configuration, R3GAN employed a 20 % burn-in and a rapid γ decay (from 150 → 15). This schedule worked well for large datasets, where tens of thousands of samples provide low-variance gradients, allowing the discriminator to establish a stable decision boundary early without overfitting. A short burn-in accelerates convergence and reduces training cost. However, for small datasets (e.g., only 3.6 k images), statistical noise is amplified and batch diversity is limited. As a result, the discriminator tends to memorize the training samples within the first 20 % of training and becomes overly confident. If γ is then rapidly reduced—thus weakening the regularization—the discriminator overwhelms the generator, producing step-like FID oscillations and late-stage rebounds.

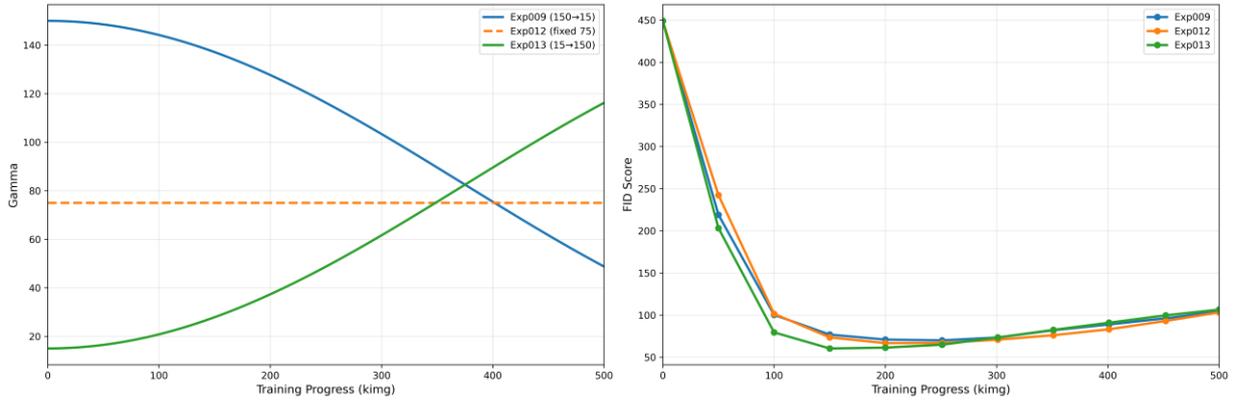

Fig. 4. Comparison of different γ scheduling strategies (exp009, exp012, and exp013). (a) γ variation during training. (b) Corresponding FID score trajectories.

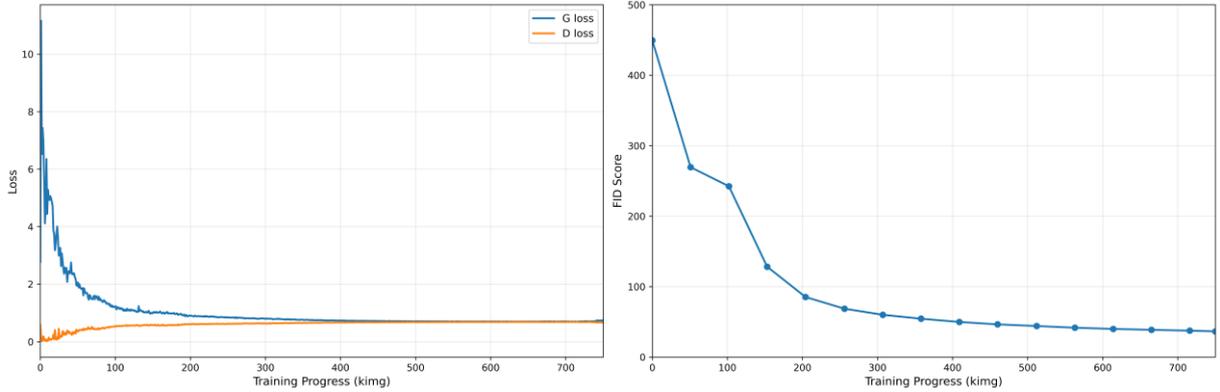

Fig. 5. Final training performance of exp017. (a) Generator and discriminator loss curves. (b) Corresponding FID score progression over 750 kimg.

Re-examining the R1/R2 regularization design reveals that its purpose is to suppress the discriminator's gradient magnitude, preventing it from driving the generator away from equilibrium. Yet as training progresses, the generator should gradually approach equilibrium, implying that γ should increase rather than decrease in the later phase. Moreover, small datasets are prone to overfitting near the end of training and thus require stronger regularization support from γ.

To verify this hypothesis, we tested two reversed γ schedules: a **fixed γ = 75** setting (**exp 012**) and a **increasing schedule 15 → 150** (**exp 013**). Compared to exp009, both configurations reduce regularization strength in the early stage and strengthen it later, while the latter applies a broader and more aggressive adjustment. Exp 012 surpassed the previous best (exp 009) with a minimum FID of **66.7**, and exp 013 further improved performance to **60.0**. When comparing the three strategies—**decreasing γ**, **fixed γ**, and **increasing γ**—(see Fig. 4), the increasing-γ setup (exp 013) achieved a noticeable lead in the early phase and comparable stability in the later phase. These results support the small-to-large γ strategy, highlighting the distinct training requirements between small and large datasets.

Consequently, all subsequent experiments adopted the increasing γ schedule as the default configuration.

### D. Lowering the γ Range and Extending Training

As shown in Fig. 4, lower γ values in the early phase consistently produced better FID scores. This observation suggests that, for small datasets of this scale, a generally weaker regularization may help the generator retain richer gradient information and avoid early over-constraining. To validate this, we reduced the overall γ range by half—**7 → 75**—and increased the late-stage augmentation probability to counter potential overfitting. This configuration was tested in **exp 014**.

Exp 014 achieved a new record FID of 57, and for the first time, the FID curve showed a steady downward trend until the end of training. The loss curves also exhibited no divergence compared to exp013, indicating that the original R3GAN γ range, which was optimized for large datasets, was indeed too high for this smaller domain. Since FID was still decreasing at the end of 500 kimg, subsequent experiments extended the total training length to **750 kimg** and further reduced γ to explore its lower limit.

Through this process, the final configuration, exp017 (γ = 5 → 40), produced exceptionally smooth loss curves and reached an FID of **36** (see Fig. 5), demonstrating the strong potential of this low-γ strategy for small datasets. Further reduction of γ did not yield any noticeable improvement.

From the baseline exp 006, which strictly followed the original FFHQ-256 configuration (20 % burn-in, γ = 150 → 15), the training initially suffered from discriminator dominance, unstable adversarial balance, and a plateaued FID around 259. Through a series of targeted adjustments—extending the burn-in period, introducing delayed-decay schedules, reversing the γ trend, and finally lowering the overall γ range with extended training—the model achieved steady improvement in both stability and generative quality. The final **exp 017** configuration (burn-in = 150 %, $\gamma = 5 \rightarrow 40$, $p_{aug} = 0 \rightarrow 0.6$, 750 kimg) reached an **FID of 36**, marking an **86 % reduction** relative to the baseline and completely eliminating the oscillatory FID patterns observed in earlier exps.

These results confirm that, compared to similar large-scale datasets, small-scale training benefits more from a longer burn-in phase, delayed and increasing $\gamma$ scheduling, and a reduced overall regularization magnitude, which are key to maintaining stable adversarial dynamics and achieving sustained improvements in generative performance.

## IV. EXPERIMENTAL RESULTS

### A. Qualitative Result

Before presenting the classification results, we first provide a qualitative demonstration of the images generated by the final configuration (exp017, $\gamma = 5 \rightarrow 40$). Fig. 6 illustrates representative embryo samples synthesized under this configuration, showing that the model successfully produces realistic and morphologically consistent embryo images across all training classes. The examples are included solely for visual reference to verify the perceptual fidelity and diversity of the generator before downstream analysis.

### B. Quantitative Comparison of Classification Results

To evaluate the effect of dataset balancing on model performance, we employed a BLIP-based [21] classifier. The BLIP (Bootstrapped Language–Image Pretraining) framework integrates a Vision Transformer (ViT) [22] as the visual encoder and a BERT-based text encoder [23] within a shared embedding space, enabling effective cross-modal alignment between visual and linguistic representations. This characteristic aligns well with the clinical workflow of embryologists, who typically examine embryo morphology and then articulate their assessments through descriptive terms. By leveraging BLIP's inherent capability to associate images with textual semantics, our classifier can better simulate this interpretative process, potentially improving stage discrimination in subtle or transient developmental phases.

In this experiment, the complete dataset comprised 4,413, 1,664, and 4,550 images corresponding to the two-cell (t2), three-cell (t3), and four-cell (t4) developmental stages, respectively. The data were divided into training, validation, and test subsets following a 70%/10%/20% split ratio. This resulted in 3,090, 1,165, and 3,185 images in the training set for t2, t3, and t4, respectively. Among these, the t3 stage was notably underrepresented—its sample size was less than half that of the other two stages—making it the minority class in the training data. To address this imbalance, we employed R3GAN with the optimized configuration (exp017, $\gamma = 5 \rightarrow 40$) to generate 2,000 additional t3 samples. The augmented training set thus contained approximately 3,090 (t2), 3,164 (t3), and 3,185 (t4) images, achieving a near-balanced distribution across classes. It is important to note that both the validation and test sets were kept identical before and after augmentation to ensure a fair comparison. None of the generated images were included in these sets, ensuring that all evaluation results were based on unseen real data.

The quantitative comparison before and after dataset balancing is summarized in TABLE I and TABLE II. Before balancing, the classifier exhibited extremely poor recognition of the t3 stage, with a recall of only 0.06 and an F1-score of 0.11. This result indicates that the model failed to capture the morphological patterns of t3 and almost never classified any image as belonging to this stage. After incorporating R3GAN-generated samples for data balancing, the recall of t3 dramatically increased to 0.69, and its F1-score improved from 0.11 to 0.60, indicating that the additional synthetic samples effectively enhanced the model's ability to identify this stage.

As shown in TABLE II, the recall of t4 slightly decreased after balancing, whereas its precision increased, resulting in an **unchanged** F1-score of 0.85. This mild trade-off is likely due to the model becoming more selective in predicting t4, reducing false positives while maintaining overall accuracy. The final F1-score did not decline, indicating that the model retained its ability to distinguish t4 effectively. The t2 metrics remained consistently high in both conditions, reflecting stable recognition of early cleavage stages. Overall, dataset balancing led to a more even classification performance across all classes. In addition, the macro-averaged precision, recall, and F1-score all improved after balancing, increasing from 0.68 to 0.80 (+**17.6%**), 0.63 to 0.82 (+**30.2%**), and 0.65 to 0.81 (+**24.6%**), respectively. These results confirm that R3GAN-based augmentation improved the classifier's overall balance among classes.

Hardware: CPU/ AMD EPYC 7763 64-Core Processor with NVIDIA GPU/H100 and RAM/64 GB.

Software: Linux Ubuntu 24.04.1 LTS, Python 3.12.3, Cuda 12.4, cuDNN 9.1.0, PyTorch 2.6.0.

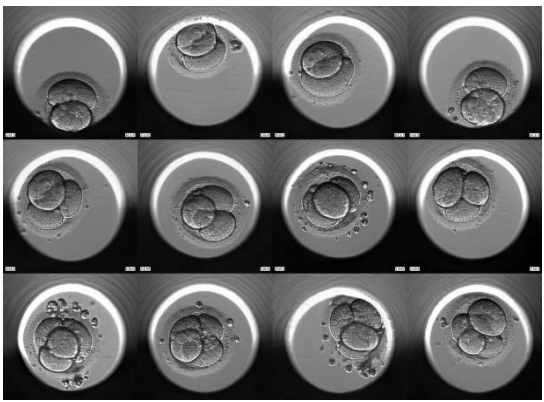

Fig. 6. Qualitative result of the final (exp017) configuration. The first, second, and third rows correspond to the t2, t3, and t4 developmental stages, respectively.

TABLE I. CLASSIFICATION METRICS BEFORE DATA BALANCING

| Category | Precision↑ | Recall↑ | F1-score↑ |
|---|---|---|---|
| t2 | 0.97 | 0.99 | 0.98 |
| t3 | 0.80 | 0.06 | **0.11** |
| t4 | 0.74 | 0.99 | 0.85 |

TABLE II. CLASSIFICATION METRICS AFTER DATA BALANCING

| Category | Precision↑ | Recall↑ | F1-score↑ |
|---|---|---|---|
| t2 | 0.96 | 0.96 | 0.96 |
| t3 | 0.52 | 0.69 | **0.60** |
| t4 | 0.91 | 0.80 | 0.85 |

## V. CONCLUSION

This study addressed the challenge of limited and imbalanced medical image datasets, a persistent obstacle in developing reliable machine learning systems for clinical applications. Using human embryo time-lapse imaging (TLI) as an example, we explored how generative adversarial networks (GANs) can be effectively trained under small-data conditions to produce realistic and diagnostically meaningful images.

Building upon the R3GAN framework, we conducted a series of controlled experiments to analyze the influence of key hyperparameters, such as burn-in phase length and $\gamma$ regularization schedules. Through these experiments, we established a set of effective training strategies and, based on them, designed an optimized R3GAN configuration specifically tailored for 256×256-resolution datasets. This configuration (exp017), characterized by a full burn-in phase and a low, gradually increasing $\gamma$ range (5 → 40), achieved stable convergence and produced high-quality generated images suitable for downstream medical image analysis.

The generated samples successfully captured the morphological characteristics of different developmental stages (*t2*, *t3*, and *t4*), demonstrating the capability of the optimized R3GAN to augment small embryo datasets. By incorporating approximately 2000 generated *t3* images into the training set, we balanced the dataset and retrained a BLIP-based classifier. The results showed that *t3* recognition significantly improved—recall increased from **0.06 to 0.69**, and F1-score from **0.11 to 0.60**—while maintaining comparable performance in other classes. These findings verify that the proposed R3GAN training strategy can effectively mitigate data scarcity, enhance classification balance, and improve model robustness in small medical imaging datasets.

Overall, this work provides a practical reference for adapting GAN-based generative models to limited medical data. Future work will extend this approach to larger and more diverse embryo datasets, and explore multi-stage generative frameworks that jointly optimize image synthesis and clinical classification performance.